\documentclass{PoS}

\usepackage{wrapfig}
\usepackage{subfigure}

\title{First results from 18-22cm VLBA polarisation observations of the MOJAVE-I AGNs}

\ShortTitle{First results from 18-22cm VLBA polarisation observations of the MOJAVE-I AGNs}

\author{\speaker{Colm Coughlan}\\
       Department of Physics, University College Cork, Ireland\\
       E-mail: \email{colmcoughlanirl@gmail.com}
}
\author{Ronan Murphy\\
       Department of Physics, University College Cork, Ireland\\
       E-mail: \email{108461490@umail.ucc.ie}
}
\author{Kyle Mc Enery\\
       Department of Physics, University College Cork, Ireland\\
       E-mail: \email{k.r.mcenery@gmail.com}
}
\author{Heather Patrick\\
       Department of Physics, University College Cork, Ireland\\
       E-mail: \email{h.patrick@umail.ucc.ie}
}
\author{Redmond Hallahan\\
       Department of Physics, University College Cork, Ireland\\
       E-mail: \email{drhallahan@gmail.com}
}
\author{Denise Gabuzda\\
       Department of Physics, University College Cork, Ireland\\
       E-mail: \email{d.gabuzda@ucc.ie}
}

\abstract{We are in the process of obtaining VLBA polarisation data for the 135 MOJAVE-I Active Galactic Nuclei at four wavelengths in the 18-22cm band.
 These observations will enable studies of the evolution of the intensity and magnetic-field structures of these AGN jets as they propagate from parsec to
 kiloparsec scales, as well as studies of the thermal plasma present in the vicinity of the jets on these scales, manifest via Faraday rotation.
 A wide range of other multi-wavelength studies can also be carried out using these data. Preliminary results for selected sources from the first 3 of 9
 observing sessions will be presented. We aim to have 18-cm intensity and polarisation images available via the MOJAVE website within 18 months after the
 last observing session.}

\FullConference{10th European VLBI Network Symposium and EVN Users Meeting: VLBI and the new generation of radio arrays \\
                September 20-24, 2010\\
                Manchester, UK}

\begin{document}

\section{Studying the Jets of AGN on scales of 10-100 pc}

Active Galactic Nuclei (AGN) are compact regions at the centres of galaxies where there is believed to be a supermassive black hole which
attracts mass to it via its gravitational pull. The mass attracted to the black hole forms an accretion disk, and as mass spirals
from the accretion disk to the black hole, it heats up and begins to emit large amounts of high-energy radiation.
It is in part this intense emission from the accretion disk that makes the AGN considered to be `active'.\\

Much of the gas from the accretion disk spirals in towards the black hole until it passes the event horizon, after which no further emission can be
 detected. However, for reasons not fully understood, but which are thought to be connected with strong magnetic fields generated by the relativistic
 plasma in the accretion disk, in about 10\% of AGN some of the gas falling into the black hole is instead accelerated to relativistic speeds and sent
 flying away from the AGN along collimated relativistic `jets', presumably along the rotational axis of the system. These jets can be detected
 because of the large amounts of radio synchrotron radiation they emit. If the jet is orientated
towards the Earth then relativistic beaming boosts the electromagnetic flux from the jet received on Earth, making the jet more prominent.\\

The polarisation of the radiation detected from the jet is of particular interest because it provides information about the underlying synchrotron magnetic field: in optically thick/thin regions the emitted polarisation is parallel/perpendicular to this magnetic field. Changes in the polarisation angle of the radiation with
frequency due to Faraday rotation can give information about both the magnetic field and thermal electron density in the immediate vicinity of the jet. Multiwavelength VLBA observations at 18-22cm probe the jet on scales from parsecs to 100s of parsecs, making them a useful tool
to study Faraday rotation. Other jet physics that can be studied using such observations include jet collimation, bending and interaction with the surrounding medium and the evolution and morphology of the magnetic field along the jet. Specific proposed magnetic field structures, such as helical jet magnetic fields (expected due to the rotation of central accretion disk, jet outflow and shocks), can also be investigated. \\

\section{The Experiment : BG 196}

We are obtaining simultaneous VLBA observations at 1358, 1430, 1493 and 1665 MHz (22.1, 21.0, 20.1 and 18.0 cm) of the 135 AGN making up the MOJAVE-1 sample~\cite{lister}. The first observations were taken in early 2010 and the observations and processing are expected to be finished in 2011-2012. These observations are designed for intensity and polarisation studies of AGN jets on scales out to 100s of parsecs. The spectral index and Faraday rotation distributions can also be studied. The first 3 epochs (A,B and C) have been calibrated, and some preliminary results are presented here.

\section{Calibration and Imaging}

\begin{figure}
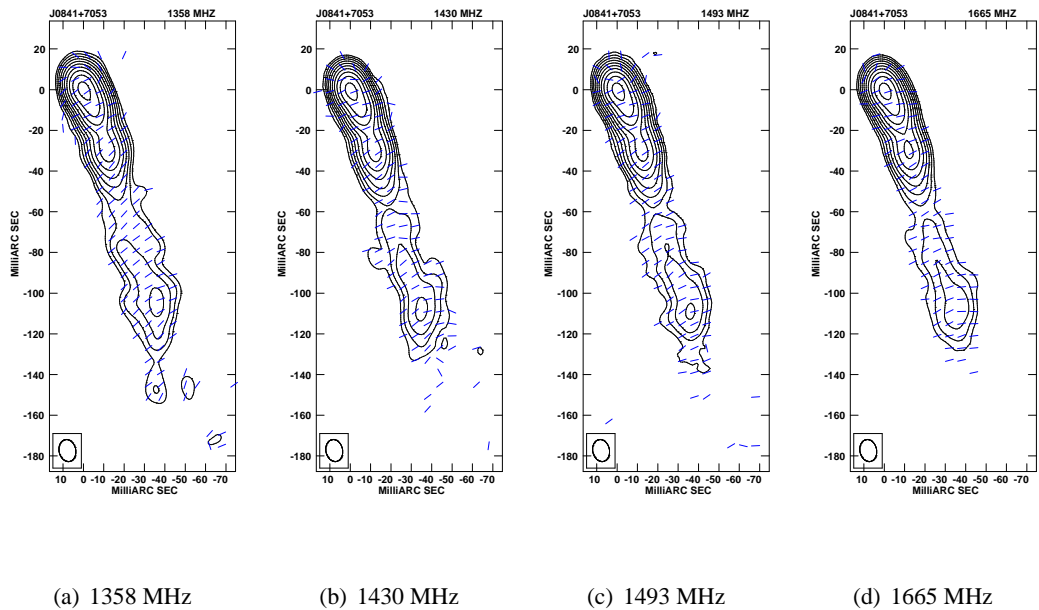

\label{cal_compare}
\begin{center}
\caption{J0841+7053 (Epoch C) with EVPA corrections applied at all four frequencies.}
\subfigure[1358 MHz]
{
  \includegraphics[scale=0.3]{J0841+7053_1_POL_UNCORR_EVPACORR_MAN.PS}
}
\subfigure[1430 MHz]
{
  \includegraphics[scale=0.3]{J0841+7053_2_POL_UNCORR_EVPACORR_MAN.PS}
}
\subfigure[1493 MHz]
{
  \includegraphics[scale=0.3]{J0841+7053_3_POL_UNCORR_EVPACORR_MAN.PS}
}
\subfigure[1665 MHz]
{
  \includegraphics[scale=0.3]{J0841+7053_4_POL_UNCORR_EVPACORR_MAN.PS}
}
\end{center}
\end{figure}

The preliminary calibration and polarisation (D-term) calibration were done in the Astronomical Image Processing System (AIPS) using standard techniques. In all cases, Los Alamos was used as the reference antenna. The polarisation angles for epochs A and B were calibrated using integrated (VLA) observations of compact polarised sources. However as no electric vector position angle (EVPA) calibrator was included in epoch C, this epoch was instead calibrated using a previous VLBI observation. Table 1 gives the resulting EVPA corrections for each frequency for each epoch. For comparison we also present the EVPA correction for the same frequencies for the observations of Gabuzda and Hallahan \cite{2004}, obtained in January 2004, which were reduced using the same reference antenna. The EVPA corrections are sometimes, but not always, remarkably stable over as long as 6 years. These EVPA corrections were tested to ensure that they resulted in self-consistant images for a range of sources. One example is shown in Figure 1, which shows the calibrated polarisation images for J0841+7053 observed at Epoch C: the polarisation structures are clearly consistant across the four frequencies. Semi-automated scripts were written in AIPS to speed up the creation of intensity, polarisation, spectral index and Faraday rotation maps.

\begin{table}

\begin{center}
  \begin{tabular}{| c | c | c | c | c | }
    \hline Frequency & Jan 2004 & Feb 2010 & Mar 2010 & May 2010 \\
     MHz & BG139 & Epoch A & Epoch B & Epoch C \\
      & $\Delta \chi^{\circ}$ & $\Delta \chi^{\circ}$ & $\Delta \chi^{\circ}$ & $\Delta \chi^{\circ}$ \\
    \hline 1358 & 128 & 130 & 91 & 132 \\
    \hline 1430 & 113 & 113 & 112 & 113 \\
    \hline 1493 & 100 & 147 & 84 & 96 \\
    \hline 1665 & 82 & 91 & 49 & 37 \\
    \hline 
  \end{tabular}
\caption{EVPA corrections for epochs A, B and C and the earlier experiment BG 139.}
\end{center}
\label{cal_table}

\end{table}

\section{Results}

\begin{wrapfigure}{r}{8cm}
\caption{3C 120.}
\subfigure[Faraday Corrected Polarisation Map at 1358 MHz]
{
  \includegraphics[scale=0.3,angle=-90]{3C120_POL_PAPER.PS}
  \label{t3c_fcpol}
}
\subfigure[Rotation Measure Map]
{
  \includegraphics[scale=0.3,angle=-90]{3C120_RM_PAPER.PS}
  \label{t3c_rm}
}
\subfigure[RM of 3C 120 across the jet. About 40 mas from the core.]
{
  \includegraphics[scale=0.3]{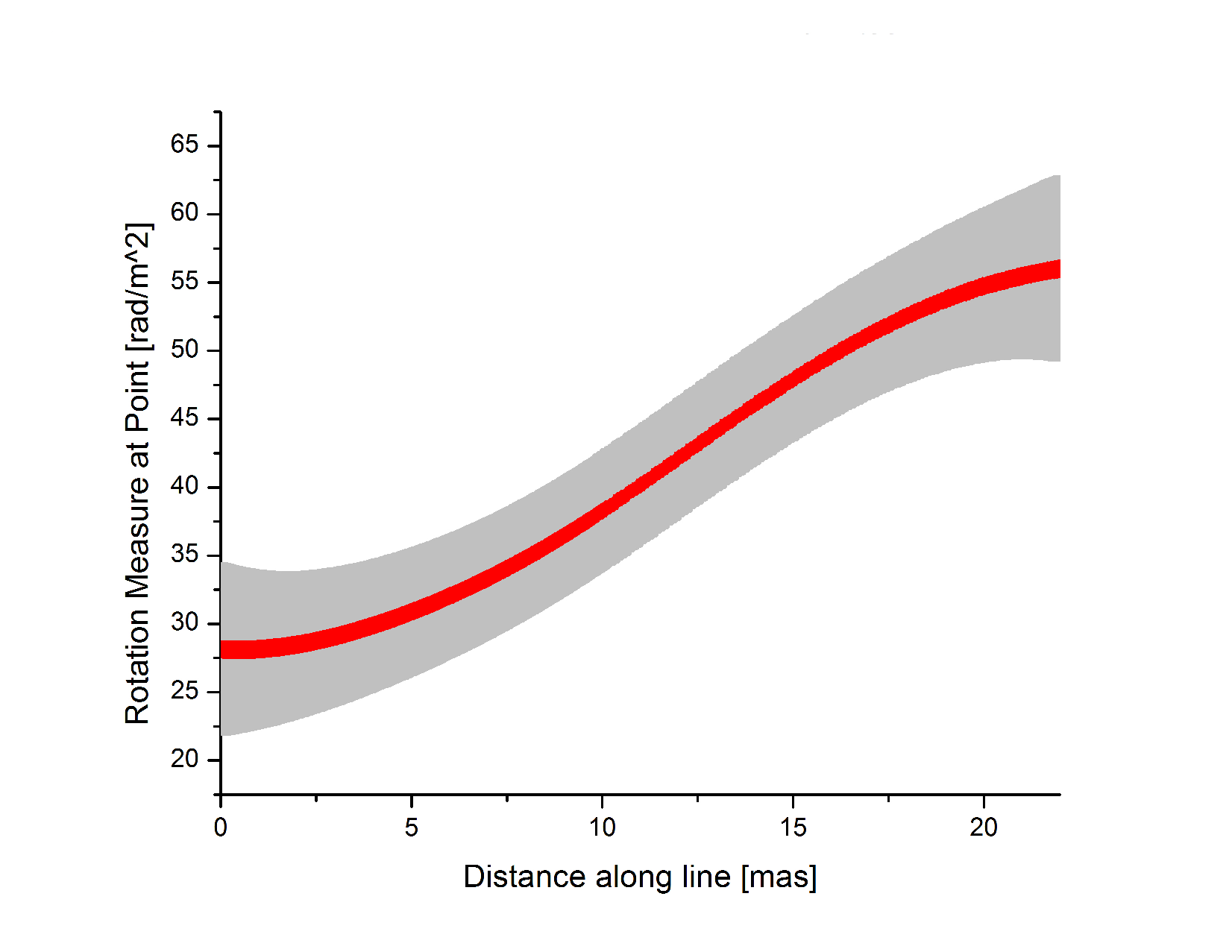}
  \label{t3c_chi2}
}
\end{wrapfigure}

Intensity, polarisation, spectral index and Faraday rotation maps have been obtained for the first three epochs. The expected wavelength dependence for Faraday rotation is

\[ \chi_{obs} = \chi_{o} + RM \lambda^{2}\]
\[ RM \propto \int n_{e} \vec{B} \cdot d \vec{l}\]
\noindent
where $n_{e}$ is the electron density, $\vec{B}$ is the magnetic field and the integral is taken along the line of sight. The amount and direction of rotation are thus determined by the electron density and the line-of-sight magnetic field. Results for two individual sources are presented below.

\subsection{3C 120}

The well known AGN 3C 120 has an extended jet with complex polarisation structure. The polarisation map of 3C 120 at 1358 Mhz presented in Figure 2(a) has been corrected for Faraday rotation. The rotation measure distribution is shown in Figure 2(b). A tendancy for the RM to be higher along the upper edge of the jet of 3C 120 is visible; one possible origin for this transverse gradient in Faraday rotation is a helical magnetic field carried by the jet: the Faraday rotation gradient is due to a systematic change in the line-of-sight magnetic field across the jet. Figure 2(c) shows a rotation measure slice across the jet about 40 mas from the core, which displays a clear systematic gradient. Previous evidence that the jet of 3C 120 carries a helical field has been presented by G\'omez et al \cite{gomez}.

\subsection{J0121+1149}

\begin{figure}
\centering
\caption{Unusual behaviour in J0121+1149. Figures 3(a) and 3(c) show the faraday corrected polarisation and RM maps during May 2010, Figures 3(b) and 3(d) show the same maps during January 2004. Figure 3(e) shows $\chi$ vs. $\lambda^{2}$ plots for the three regions during May 2010. The $\chi$ errors are comparable to the size of the symbols.}
\subfigure[]{
\includegraphics[trim = 10mm 30mm 1mm 5mm, clip,scale=0.35]{J0121+1149_POL_FINAL_MAN.PS}
}
\subfigure[]{
\includegraphics[trim = 10mm 30mm 1mm 5mm, clip,scale=0.35]{J0121+1149_POL_FIN_OLD_MAN.PS}
}
\subfigure[]{
\includegraphics[trim = 0mm 10mm 0mm 13mm, clip,scale=0.3]{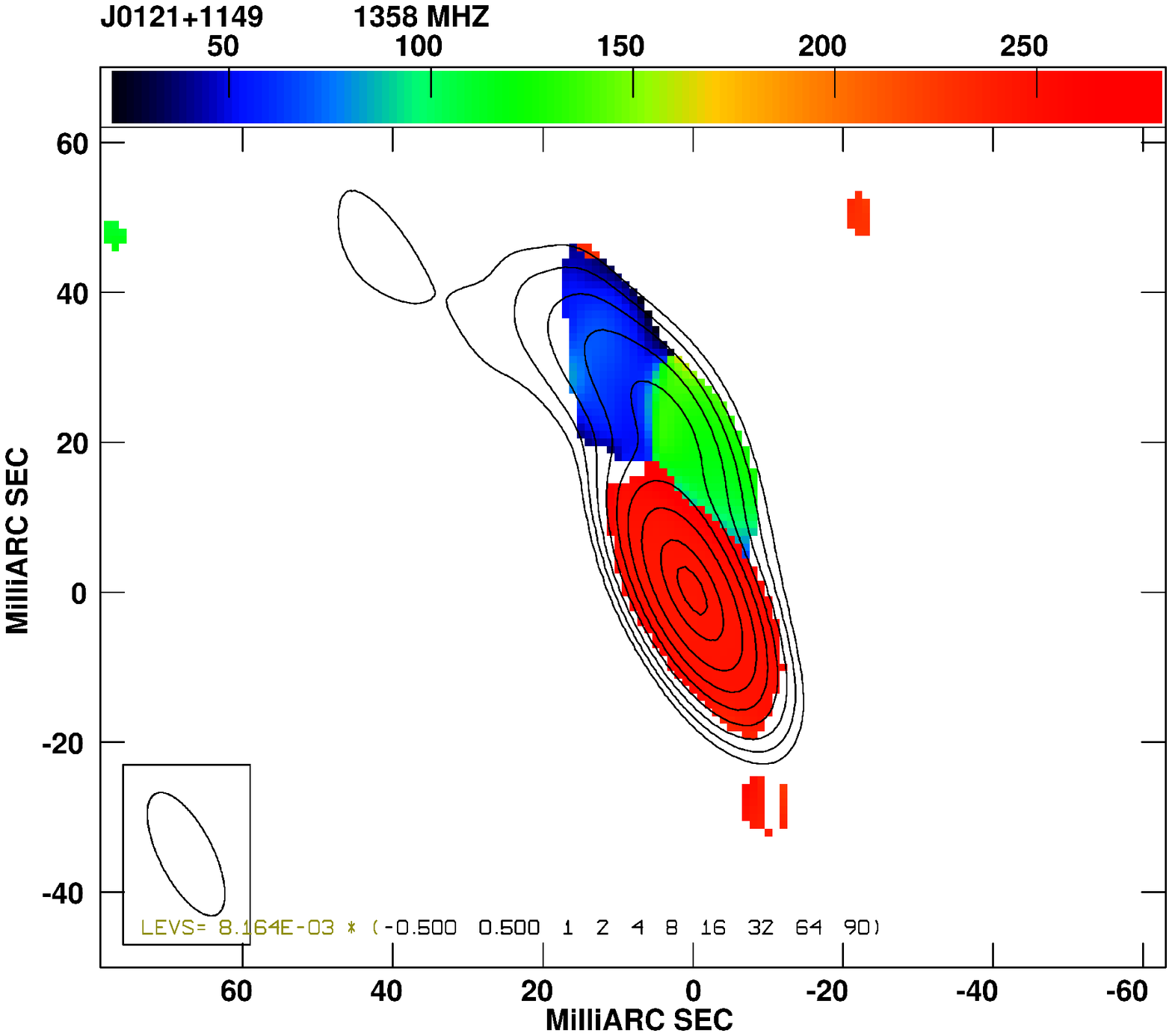}
}
\subfigure[]{
\includegraphics[trim = 0mm 20mm 0mm 10mm, clip,scale=0.3]{J0121+1149_RM_OLD_MAN.PS}
}
\subfigure[]{
\includegraphics[trim = 0mm 5mm 0mm 5mm, clip,scale=0.37]{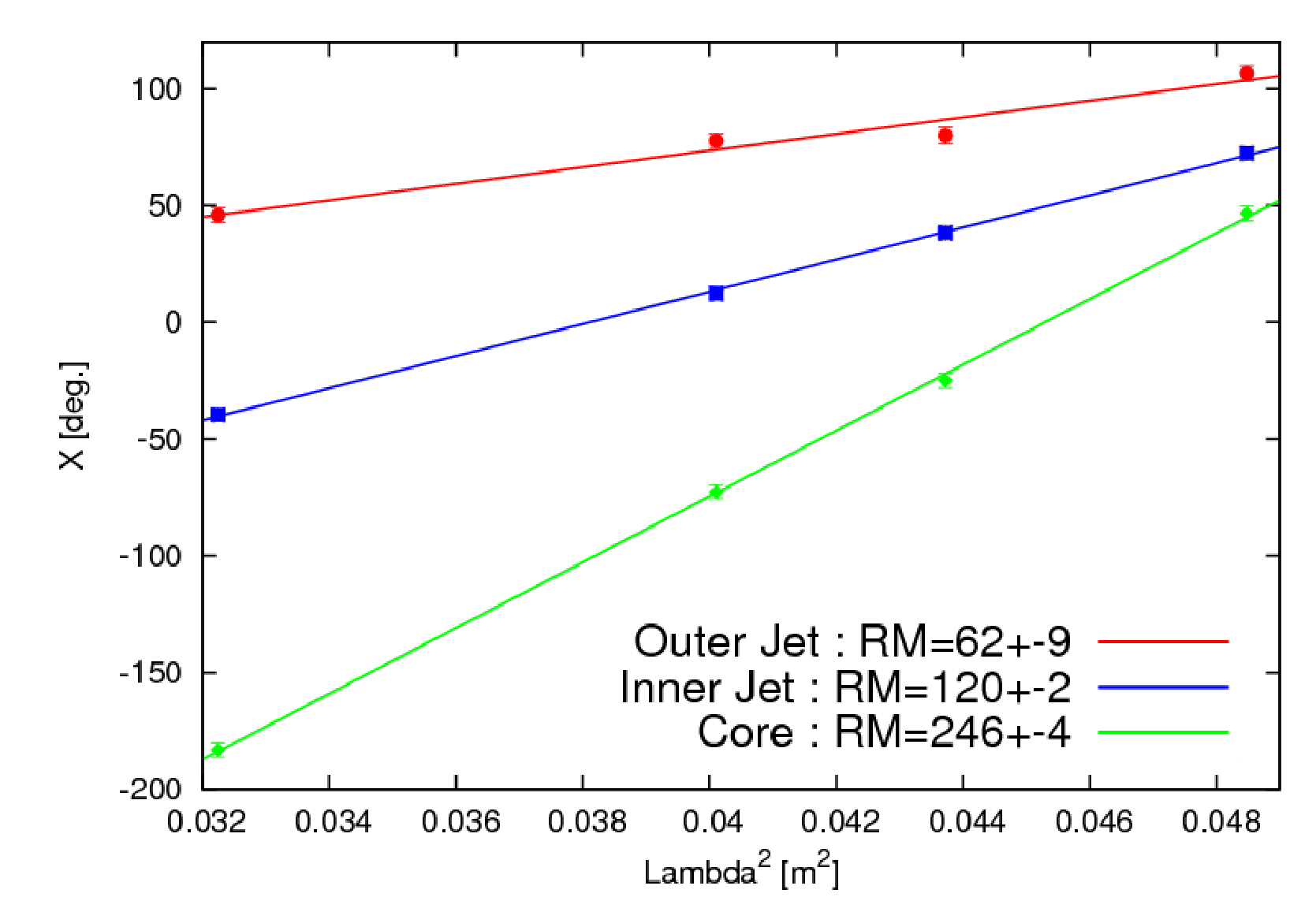}
}
\label{j0121}
\end{figure}

The Faraday rotation map of this source, shown in Figure~\ref{j0121}, displays 3 distinct regions with different characteristic RM values: the core, inner jet and outer jet. After correcting for this Faraday rotation, the magnetic field structure is complex, but appears to become longitudinal in the outer jet. Figure~\ref{j0121} also shows a comparison of the polarisation images obtained at epoch C and in January 2004. The three distinct regions of the Faraday rotation are present at both epochs. The polarisation structure in the core and inner jet changed between the two epochs, while that in the outer jet remained constant. The $\chi$ vs. $\lambda^{2}$ plots in Figure~\ref{j0121} clearly show the linear dependence expected for Faraday rotation, with the RM decreasing markedly with distance from the core.

\section{Conclusions}

The images presented here show just two examples of the suitability of 18-22 cm VLBA observations for studying AGN jets on scales out to 10s or 100s of parsecs. We have worked out a scheme for calibrating and imaging the data for BG196 in a timely fashion, including the development of semi-automated scripts. We plan to provide calibrated data and images to the community within 18 months after the last observations, via the MOJAVE website \cite{lister}.

\section*{Acknowlegements}

This work was supported by the Irish Research Council for Science, Engineering and Technology (IRCSET).

\end{document}